\documentclass[11pt]{article}

\usepackage[preprint]{acl}

\usepackage{times}
\usepackage{latexsym}

\usepackage[T1]{fontenc}

\usepackage[utf8]{inputenc}
\usepackage{graphicx}
\usepackage{hyperref}
\usepackage{fontawesome5}
\usepackage{microtype}

\usepackage{inconsolata}

\usepackage{graphicx}
\usepackage{booktabs}
\usepackage{multirow}
\usepackage{siunitx}
\title{\method: Graph-Based Profiling for Cold-Start LLM Routing}

\author{
Jingjun Xu$^{1*}$,
Hongji Pu$^{1*}$,
Tao Feng$^{1*}$,
Haozhen Zhang$^{2*}$, \\
Jiaxuan You$^{1\dagger}$,
Ge Liu$^{1\dagger}$ \\
$^1$University of Illinois Urbana-Champaign \\
$^2$Nanyang Technological University \\
\texttt{\{jingjunx,hongjip2,taofeng2\}@illinois.edu} \\
\texttt{wazhz14@gmail.com} \\
{\small $^*$Equal contribution. \quad $^\dagger$Corresponding author.} \\
\vspace{-0.7em} \\
{\large
\href{https://github.com/ulab-uiuc/RouteProfile}{\faGithub\ ulab-uiuc/RouteProfile}
\quad
\href{https://huggingface.co/collections/ulab-ai/routeprofile}{
\includegraphics[height=0.9em]{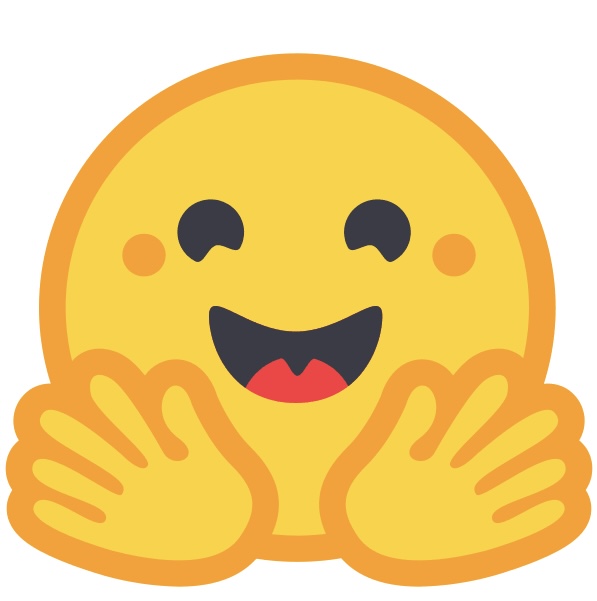}\ Hugging Face Collection}
}
}
\usepackage{threeparttable} 
\usepackage{booktabs}
\usepackage{amsmath}
\usepackage{enumitem}
\usepackage{xspace}
\usepackage{makecell}   
\usepackage{arydshln}   
\usepackage{multirow}   
\usepackage{pifont}     
\newcommand{\cmark}{\ding{51}}
\newcommand{\method}{\texttt{RouteProfile}\xspace}
\newcommand{\dashmidrule}{%
  \noalign{\vskip\aboverulesep}
  \hdashline
  \noalign{\vskip\belowrulesep}
}

\begin{document}
\maketitle

\begin{abstract}
LLM routing is increasingly important for selecting suitable models under diverse user needs and deployment constraints, but its practical effectiveness depends on continual adaptation to emerging queries and newly released models.
New-LLM integration is particularly challenging, as newly released models lack the query-response-reward interactions required for router training and cannot be profiled as directly as new queries via semantic embeddings. 
Existing profiles are limited: LLM-generated descriptions are often coarse, while interaction-based embeddings are costly to construct.
To address this problem, we propose \method, a graph-based profiling framework that constructs LLM profiles from public signals in technical reports or model cards, including model family, model description, reported benchmark scores, and benchmark domains. 
\method organizes these heterogeneous signals into a graph and studies profile construction along four dimensions: \textit{organizational form}, \textit{representation type}, \textit{aggregation depth}, and \textit{learning configuration}. 
We evaluate \method in training-free cold-start routing and new-LLM integration settings.
Experiments show that: 
(1) structured profiles outperform flat baselines in training-free cold-start routing; 
(2) model family metadata is more reliable than benchmark domain information; and 
(3) effective new-LLM integration requires profile--router co-design. 
Overall, our findings highlight the importance of profile design for enabling routing systems to adapt to the evolving model ecosystem.
\end{abstract}

\section{Introduction}

\begin{figure*}[t]
    \centering
    \includegraphics[width=0.85\textwidth]{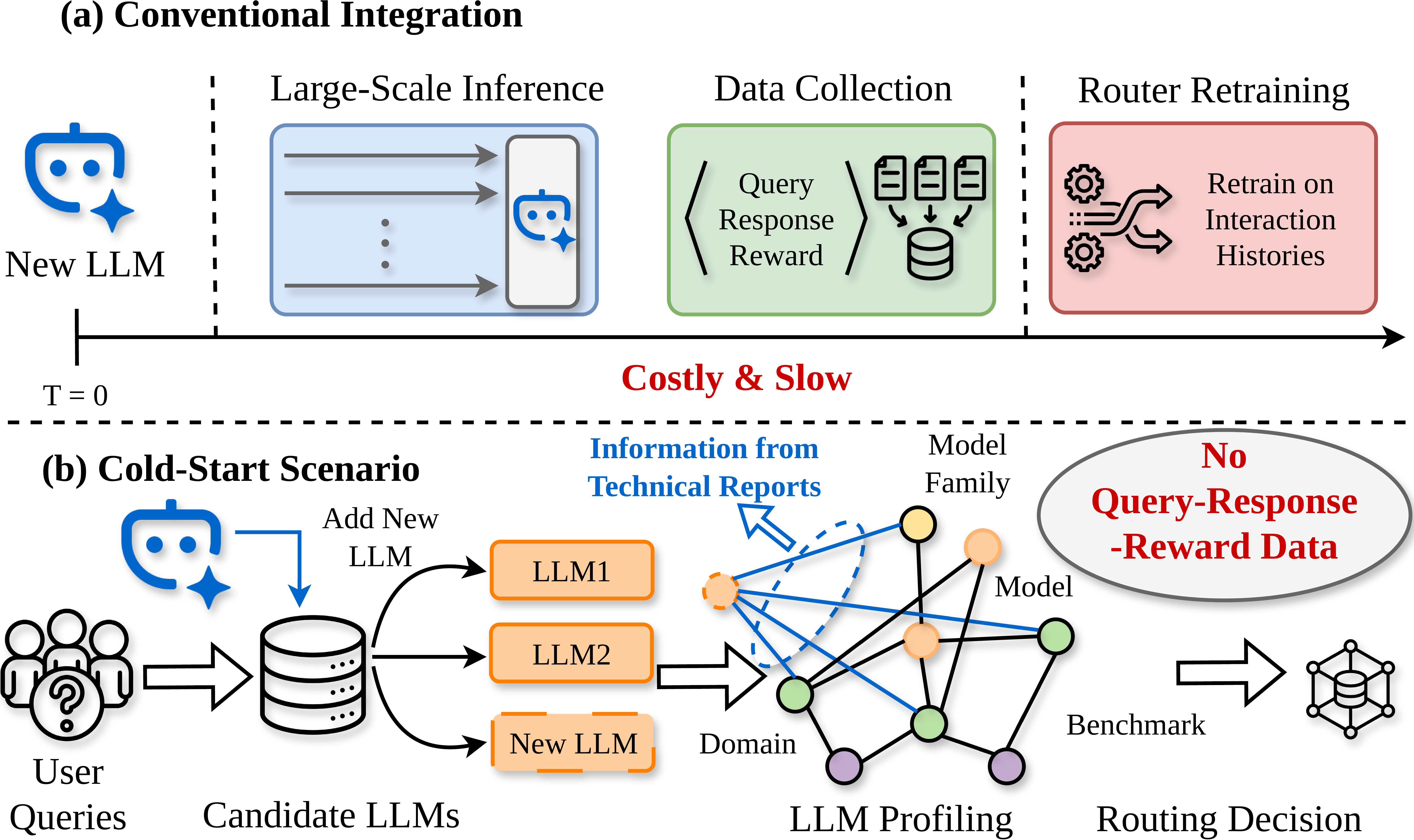}
    \caption{\textbf{Cold-start LLM Routing.}
(A) Conventional routing incurs high latency and cost for new-model integration due to large-scale inference, data collection, and router retraining; (B) Cold-start LLM routing profiles LLMs from public signals, including model family metadata, model description, reported benchmark scores, and benchmark domain information.}
    \label{fig:intro}
\end{figure*}

As the large language model (LLM) ecosystem expands, individual models exhibit heterogeneous capabilities across queries, benchmarks, and domains. 
This heterogeneity motivates LLM routing, which selects the most suitable model for each query under diverse user needs and deployment constraints~\citep{chen2023frugalgpt}. 
However, the practical effectiveness of routing systems depends on their ability to continually adapt to emerging queries and newly released models. 
New-LLM integration is particularly challenging: newly released models lack the query--response--reward interactions typically required for router training, and unlike new queries, they cannot be directly profiled through semantic embeddings. 
Existing routing systems, therefore, require large-scale inference, query-level data collection, and router retraining to incorporate new models, making adaptation costly and slow as model releases accelerate, as illustrated in Fig.~\ref{fig:intro}.
This motivates cold-start LLM routing, where a target LLM must be routed without query-response-reward data, analogous to cold-start recommendation for new items without interaction histories~\citep{schein2002methods}. 
Our key observation is that newly released LLMs often come with public signals from technical reports or model cards, which may support cold-start profiling despite the absence of query-response-reward interactions. 
We therefore ask: \textit{Can LLMs be effectively profiled using such coarse public signals to support cold-start routing?}

Constructing model profiles under the cold-start constraint is challenging because newly released LLMs lack the query--response--reward interactions typically used for router training. 
Profiles must therefore be inferred from coarse public signals in technical reports or model cards, including model family, model descriptions, reported benchmark scores, and benchmark domains. 
Although accessible, these signals are sparse, heterogeneous, and only partially comparable: models are evaluated on different benchmark subsets, while textual descriptions, numerical scores, and relational metadata encode different types of information.

Existing LLM profile designs remain inadequate for cold-start routing. 
Index-based profiles represent each model as a one-hot vector~\citep{zheng2023judging}, providing little semantic information for generalizing to unseen queries or new models. 
LLM-generated profiles describe candidate models using a strong LLM~\citep{feng2025graphrouter, zhang2025router}, but are often coarse, knowledge-limited, and incomplete. 
Benchmark-level summaries~\citep{benchmark_sum} are cheaper to obtain, but discard structured relationships among models, benchmarks, and domains.

These limitations suggest that effective cold-start routing requires profiles that are both informative and structurally grounded. 
We therefore propose \method, a graph-based profiling framework that organizes heterogeneous public signals into a structured graph and analyzes profile construction along four dimensions: \textit{organizational form}, \textit{representation type}, \textit{aggregation depth}, and \textit{learning configuration}. 
\textbf{Organizational form} specifies whether profiles leverage graph structure. 
\textbf{Representation type} determines whether profiles are constructed through textual summaries or dense embedding computations. 
\textbf{Aggregation depth} controls how far information propagates over the graph. 
\textbf{Learning configuration} indicates whether aggregation is training-free or optimized through learning. 
Rather than enumerating all possible design variants, we aim to identify which profile design choices most critically affect routing performance under the cold-start constraints.


We systematically evaluate \method to understand how profile design affects cold-start LLM routing.
Experiments cover two complementary scenarios: \textbf{training-free cold-start routing} with SimRouter, where no candidate LLM has query-response-reward interactions, and \textbf{new-LLM integration} with MLPRouter~\citep{hu2024routerbench} and GraphRouter~\citep{feng2025graphrouter}, where a newly released LLM is added to an existing router using only its public profile, without retraining. 
In both settings, all profiles are constructed exclusively from public signals.
Our evaluation yields three key findings: 
\textbf{(1)} profile structure (organizational form and aggregation depth) substantially improves training-free cold-start routing; 
\textbf{(2)} model family metadata is a more reliable public signal than benchmark domain information; 
\textbf{(3)} effective new-LLM integration depends critically on profile–router co-design (representation type and learning configuration).
Overall, our results highlight profile design as a critical component for adapting routing systems to the expanding LLM ecosystem.

\section{Related Work}

\textbf{LLM Routing.} 
Recent work formulates multi-LLM routing as an inference-time decision problem, assigning each query to a model under quality, cost, or latency constraints \citep{ding2024hybridllm, ong2025routellm, chen2023frugalgpt}. Existing methods mainly focus on router design, including preference-trained, reward-guided, contrastive, and graph-based routing \citep{zhang2025router, ong2025routellm, chen2024routerdc, feng2025graphrouter, sakota2024flyswat}. Some methods also use model-side signals such as benchmark statistics, metadata, or structured benchmark--query--model relations \citep{ong2025routellm, chen2024routerdc, feng2025graphrouter}, but typically treat them as auxiliary inputs rather than a standalone design problem. In contrast, we study LLM profile design and its effect across routers.

\textbf{LLM Profiling.} 
Prior work studies explicit profiling of model capabilities. QualEval~\citep{murahari2024qualeval} derives natural-language capability groups for diagnosis, Skill-Slices~\citep{moayeri2024unearthing} recovers latent skills to reveal trade-offs hidden by aggregate benchmark scores, and EvalTree~\citep{zeng2025evaltree} organizes model weaknesses through capability trees. More recently, BELLA explores skill-based profiling for cost-aware LLM routing~\citep{okamoto2026trust}. However, these works mainly target evaluation, diagnosis, or a specific routing framework, rather than profile design as a general routing problem.
\begin{figure*}[t]
    \centering
    \includegraphics[width=\textwidth]{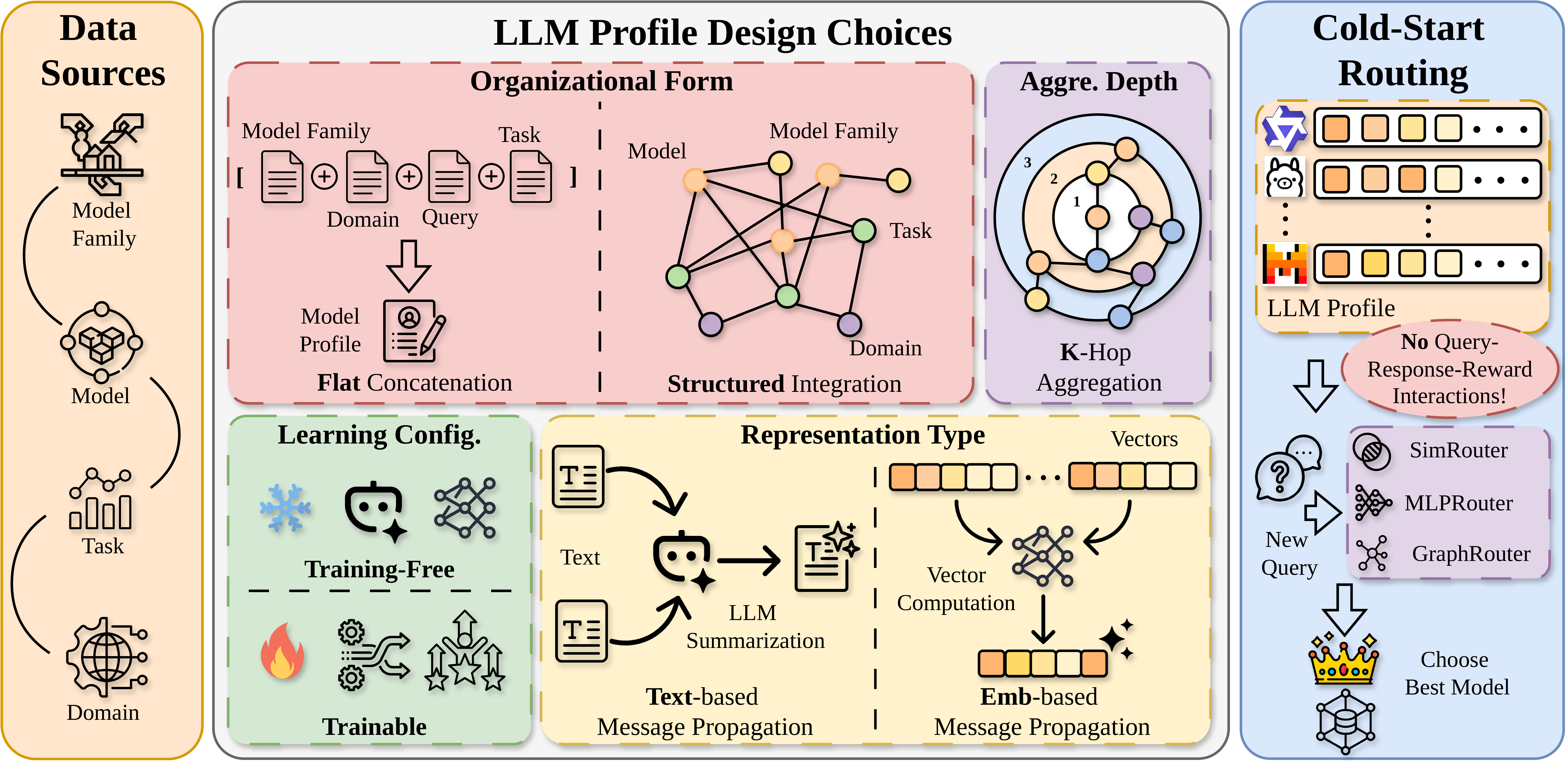}
    \caption{\textbf{Overview of the \method.} LLM profiles are constructed from coarse public signals comprising model family metadata, model description, reported benchmark scores, and benchmark domain information. The design choices are characterized along four dimensions: \textit{organizational form} (flat/structured), \textit{representation type} (text/embedding), \textit{aggregation depth} (hop $\in \{0,1,2, ...\}$), and \textit{learning configuration} (training-free/trainable). Experiments are conducted across three representative routers to evaluate how profile design choices affect cold-start routing performance across different routing settings. "Aggre." and "Config." denote Aggregation and Configuration.}
    \label{fig:design_space}
\end{figure*}

\section{Public Signals as a Heterogeneous Graph for Cold-start Profiling}

This section first introduces the public data sources used for profile construction, and then formalizes them as a heterogeneous graph for principled LLM profile definition and systematic analysis.

Under the cold-start constraint, LLM profiles must be constructed without query-response-reward data, relying only on coarse public signals such as reported benchmark scores and model metadata. 
As illustrated in Figure~\ref{fig:design_space}, we construct LLM profiles from four primary sources: \textit{model family metadata}, \textit{model description}, \textit{reported benchmark scores}, and \textit{benchmark domain information}. 
\textbf{Model family} encodes the structural prior of each model, including its architectural lineage, series, and developer, and thus provides insight into inherent capabilities.
\textbf{Model description} summarizes publicly available textual description of the model from technical reports or model cards, such as training objectives and model scale.
\textbf{Benchmark evaluation} captures the model's standardized performance in technical reports or model cards and, therefore, offers a comparable assessment of model capabilities.
\textbf{Domain coverage} characterizes the benchmark areas in which a model exhibits competence, highlighting its specialization and heterogeneity across domains.


Public data sources are sparse, heterogeneous, and only partially comparable, making direct profile construction challenging. 
To systematically integrate them, we formulate the multi-source information as a heterogeneous graph $\mathcal{G} = (\mathcal{V}, \mathcal{E})$. 
Each node $v \in \mathcal{V}$ and edge $e \in \mathcal{E}$ are assigned types through mapping functions, with node type defined by $\phi: \mathcal{V} \rightarrow \mathcal{C}$ and edge type defined by $\psi: \mathcal{E} \rightarrow \mathcal{D}$.
An edge connecting a pair of nodes is denoted as $e_{uv} = (u, v)$.
Specifically, we define 4 node types: model node $v_m$, model family node $v_f$, domain node $v_d$, benchmark node $v_b$; and 3 edge types: model-model family edge $e_{mf}$, model-benchmark edge $e_{mb}$, and benchmark-domain edge $e_{bd}$.

We then describe the features associated with nodes and edges. 
For \textbf{node features} $\mathbf{x}$, we adopt different initialization strategies given the inherent differences among node types. 
In particular, we utilize an additional LLM, such as GPT-4o-mini \citep{gpt4o}, to generate textual descriptions for model nodes, domain nodes, and benchmark nodes using tailored prompts based on publicly available information.
All generated descriptions can be found in Appendix \ref{appendix:profile_evidence_description}. 
These descriptions serve as node features in the text space and are further encoded by a pre-trained language model (PLM), such as Longformer \citep{longformer}, to obtain dense embeddings.
For \textbf{edge features} $\mathbf{r}$, only the model–benchmark edges are associated with features, which encode performance scores demonstrated on technical reports or authoritative LLM leaderboards, such as the Open LLM Leaderboard \citep{open-llm-leaderboard-v2}.

Finally, we define the LLM profile $\mathbf{p}_m$ of a model node $v_m$ as:
\begin{equation}
    \mathbf{p}_m = \hat{\mathbf{x}}_{v_m} = f(\mathcal{G})_{v_m},
\end{equation}
where $\hat{\mathbf{x}}_{v_m}$ denotes the aggregated representation of $v_m$, and $f$ is the information aggregation function over the heterogeneous graph $\mathcal{G}$.

\section{\method: Profile Design for Cold-Start LLM Routing}
\label{section_design_space}

Next, we propose \method for LLM profile design in cold-start routing.
Specifically, we focus on the design of the information aggregation function $f$.

The \method includes four key dimensions as illustrated in Figure \ref{fig:design_space}: \textit{organizational form}, \textit{representation type}, \textit{aggregation depth}, and \textit{learning configuration}.
In defining the LLM profile design choices, we follow two guiding principles:
\textbf{(1)} inclusiveness of dimensions that materially affect downstream routing performance;
\textbf{(2)} conciseness by excluding overly task-specific choices, such as particular LLMs or graph neural networks (GNNs) used for information aggregation.
Our goal is not to enumerate all possible design variants, but to show how a systematic view for understanding how different profile design choices affect cold-start routing performance.


In particular, \textbf{organizational form} specifies whether the structural information in the heterogeneous graph is leveraged during aggregation.
In a structured form, relational information is typically modeled through a GNN, whereas in a flat form, the available information is directly concatenated into plain text or a single vector.
\textbf{Representation type} determines the information fusion mechanism, which can either be textual descriptions or dense embeddings.
Textual representations are usually summarized by LLMs, whereas dense embeddings are often computed through neural networks, such as those in GNNs.
\textbf{Aggregation depth} controls the extent of information propagation within the graph, determining whether only direct neighbors or also higher-order neighborhoods contribute to the LLM profiles.
\textbf{Learning configuration} indicates whether the aggregation function $f$ is trainable.
In a trainable setting, the aggregation function $f$ can be optimized, for example, via self-supervised learning on the  graph.

Formally, we define the function $f$ as:
\begin{equation}
\mathbf{p}_m = \hat{\mathbf{x}}_{v_m} = f^{(\omega,\gamma,K,\ell)}(\mathcal{G})_{v_m},
\end{equation}
where $\omega \in \{\text{Flat}, \text{Structured}\}$ denotes the organizational form, $\gamma \in \{\text{Text}, \text{Embedding}\}$ denotes the representation type, $K \in \{0,1,2,3,4\}$ denotes the aggregation depth, and $\ell \in \{\text{Training-free}, \text{Trainable}\}$ denotes the learning configuration.

\section{Experimental Setup}


In this section, we describe the experimental setup for evaluating how design choices in LLM profiles affect cold-start routing performance.
The setup comprises two parts:
upstream profile construction, covering heterogeneous graph construction (Section \ref{section_exp_setup:evidence_graph}) and instantiated design choices for constructing LLM profiles (Section \ref{section_exp_setup:design_choice}); and downstream routing evaluation, including datasets and candidate LLMs (Section \ref{section_exp_setup:dataset_llm}), evaluation tasks (Section \ref{section_exp_setup:routing_task}), and routing methods (Section \ref{section_exp_setup:routers}).

\subsection{Heterogeneous Graph Construction}
\label{section_exp_setup:evidence_graph}

We construct the heterogeneous graph using 15 datasets spanning 4 capability domains: knowledge, reasoning, math, and coding. 
Dataset statistics are summarized in the upper portion of Table~\ref{tab:datasets}, with detailed descriptions provided in Table~\ref{tab:dataset_descriptions}. 
The graph further incorporates 25 LLMs from 5 model families to enrich relational signals across model families and task performance, of which 8 serve as candidate LLMs for downstream routing evaluation and the remainder as auxiliary nodes to improve graph connectivity and evidence diversity. 
Full statistics are reported in Table~\ref{tab:llms}, with descriptions in Table~\ref{tab:model_descriptions}.

\subsection{Instantiated Design Choices for LLM Profile Construction}
\label{section_exp_setup:design_choice}

We present concrete instantiations of the aggregation function $f^{(\gamma,\omega,K,\ell)}$, covering four representative configurations that vary along the dimensions defined in Section \ref{section_design_space}.

\textbf{Flat Aggregation ($\gamma=\text{Text},\omega=\text{Flat}, K=0,\ell=\text{Training-free}$).} 
Flat aggregation constructs the LLM profile directly in the text space without exploiting graph structure.
Data associated with $v_m$ is sampled from $\mathcal{G}$ and concatenated into a textual description:
\begin{equation}
    \mathbf{p}_m = f^{(\text{Text},\text{Flat}, 0,\text{Training-free})}(\mathcal{G})_{v_m} =\mathcal{C}\!\left(\mathcal{S}(v_m)\right),
\end{equation}
where $\mathcal{S}(v_m)$ denotes the sampled data for $v_m$, and $\mathcal{C}(\cdot)$ is a concatenation operator.

\textbf{Text-based GNN ($\gamma=\text{Text},\omega=\text{Structured}, K \in \{1, 2, 3, 4\},\ell=\text{Training-free}$).} Inspired by \cite{researchtown}, the text-based GNN performs message passing entirely in the text space.
The aggregation function updates each node $v$ by prompting an LLM to summarize the textual attributes propagated from neighborhood $\mathcal{N}(v)$.
At each propagation hop $k$, a node-type-specific prompt template $\mathcal{T}(\cdot)$ organizes the current node text with neighboring textual states and available edge features into a unified prompt $\pi_v^{(k)}$:
\begin{equation}
    \pi_v^{(k)} = \mathcal{T}\!\left(\mathbf{x}_v^{(k-1)}, \{\, (\mathbf{x}_u^{(k-1)}, \mathbf{r}_{uv}) \mid u \in \mathcal{N}(v) \,\}\right),
\end{equation}
and the updated representation is obtained by querying an LLM:
\begin{equation}
    \mathbf{x}_v^{(k)} = \mathrm{LLM}\!\left(\pi_v^{(k)}\right).
\end{equation}
The LLM profile is then defined as $\mathbf{p}_m = f^{(\text{Text},\text{Structured}, K,\text{Training-free})}(\mathcal{G})_{v_m} =\mathbf{x}_{v_m}^{(K)}$.

\textbf{Embedding-based GNN ($\gamma=\text{Emb},\omega=\text{Structured}, K \in \{1, 2, 3, 4\},\ell=\text{Training-free}$).} 
The embedding-based GNN performs feature aggregation on the heterogeneous graph at the embedding level through message passing. 
Following a simplified GCN-style propagation inspired by \cite{gwm}, node representations are updated at the embedding level:
\begin{equation}
\begin{aligned}
    &\mathbf{x}_v^{(k)} = \\
    & \sum_{u \in \mathcal{N}(v)\cup\{v\}}
    \frac{w_{uv}}
    {\sqrt{|\mathcal{N}(v)\cup\{v\}|\,|\mathcal{N}(u)\cup\{u\}|}}
    \mathbf{x}_u^{(k-1)} .
\end{aligned}
\end{equation}
where $w_{uv} = \mathbf{r}_{uv}$ if an edge feature is available, and $w_{uv} = 1$ otherwise. Then the profile is $\mathbf{p}_m = f^{(\text{Emb},\text{Structured}, K,\text{Training-free})}(\mathcal{G})_{v_m} =\mathbf{x}_{v_m}^{(K)}$. 

\textbf{Trainable GNN($\gamma=\text{Emb},\omega=\text{Structured}, K \in \{1, 2, 3, 4\},\ell=\text{Trainable}$).}
The trainable GNN extends the embedding-based GNN with a learnable aggregation optimized via a self-supervised masked reconstruction objective. 
A proportion of node and edge features is randomly masked, and the model is trained to reconstruct the masked attributes from the remaining graph context:
\begin{equation}
    \mathcal{L} = \mathcal{L}_{\mathrm{node}} + \mathcal{L}_{\mathrm{edge}},
\end{equation}
where \(\mathcal{L}_{\mathrm{node}}\) and \(\mathcal{L}_{\mathrm{edge}}\) are both implemented as mean squared error (MSE) losses. The resulting profile is $\mathbf{p}_m = f^{(\text{Emb},\text{Structured}, K,\text{Trainable})}(\mathcal{G})_{v_m} =\mathbf{x}_{v_m}^{(K)}$. 

\subsection{Downstream Datasets and Candidate LLMs}
\label{section_exp_setup:dataset_llm}

We select 12 datasets spanning math, reasoning, knowledge, and coding, sampling 50 instances per dataset for downstream evaluation. Statistics are summarized in the lower portion of Table~\ref{tab:datasets}, with detailed descriptions in Table~\ref{tab:dataset_descriptions}. 
Furthermore, routing is evaluated over a fixed candidate pool of 8 LLMs drawn from the Qwen2, Llama, Gemma2, Mistral, and Mixtral families, covering model scales from 3B to 176B parameters. Full statistics are reported in Table~\ref{tab:llms}, with detailed descriptions in Table \ref{tab:model_descriptions}.

\subsection{Cold-Start Routing Tasks and Metrics}
\label{section_exp_setup:routing_task}

We evaluate \method under two complementary settings reflecting cold-start scenarios.

\textbf{Training-free cold-start.}
In this setting, no candidate LLM has query-response-reward interactions, and routing decisions rely entirely on profiles derived from public signals. 
We adopt SimRouter because it requires no training, enabling routing without query-response-reward interactions and isolating the effect of profile design from router optimization. 
The evaluation metric is average response performance across queries, as introduced in Table~\ref{tab:datasets}.
We additionally report three reference points: Oracle (per-query best model), Single-Best (the globally best single model applied to all queries), and Random (random selection averaged over seeds 1–5).
\textbf{New-LLM integration.}
This setting evaluates whether public-signal profiles allow a \emph{fixed} trained router to generalize to a newly released LLM at inference time, without retraining. 
Specifically, we first train the router on query-response-reward interaction data from seven existing candidate LLMs and then freeze its parameters. 
We then introduce the new LLM into the candidate pool by adding it as a new node in the heterogeneous graph, initialized it using different profiling methods using only public information.
The fixed router is then applied directly to user queries over the expanded candidate pool, with no parameter updates.

\texttt{Mistral-Small-24B-Instruct-2501} is designated as the new LLM in our experiments.
Beyond average performance, we define a metric called \textit{New-LLM Correct Integration Rate} (NCIR) that jointly captures selection frequency and correctness for the new LLM:
\begin{equation}
\label{eq:cold-start-performance}
     \text{NCIR} = \frac{N_{\text{new} \wedge \text{correct}}}{N},
\end{equation}
where $N$ is the total number of queries, and $N_{\text{new} \wedge \text{correct}}$ is the number of queries both routed to and correctly answered by the new LLM.




\subsection{Routing Methods}
\label{section_exp_setup:routers}



We consider three representative routers across the two settings: SimRouter is used in the training-free cold-start setting; MLPRouter and GraphRouter are used in the new-LLM integration setting.

\textbf{SimRouter} is a similarity-based, non-parametric router that selects models by measuring similarity between the query representation and each candidate's profile.
As a training-free baseline, it directly reflects profile quality without any learned transformation, making it well-suited for evaluating profiles under the cold-start constraint.

\textbf{MLPRouter}~\citep{hu2024routerbench} is a trainable router that projects query representations and model profiles into a shared latent space via separate MLPs, ranking candidates by similarity between projected representations.
In the new-LLM integration setting, it is trained on query-response-reward interactions from old LLMs only; its ability to route queries to the new LLM therefore depends on the quality of the new LLM's public-signal profile.

\textbf{GraphRouter}~\citep{feng2025graphrouter} organizes tasks, queries, and candidate LLMs into a heterogeneous graph and applies a GNN with self-supervised learning to model their relational interactions.
Like MLPRouter, it is trained exclusively on old LLM query-response-reward data in the new-LLM integration setting, and evaluates whether graph-structured relational reasoning can further leverage public-signal profiles to generalize to newly introduced models.






\section{Experimental Results}

We present results across training-free cold-start (Tables~\ref{tab:profile_design_space_compact}–\ref{tab:source_ablation}) and new-LLM integration (Table~\ref{tab:router-performance-ncir}) settings to examine how LLM profile design choices affect cold-start routing.

\begin{table*}[t]
\centering
\small
\caption{\textbf{Routing performance under training-free cold-start across profile designs.}
Abbreviations: ``Org.'' = organizational form, ``Rep.'' = representation type, ``Aggre.'' = aggregation depth, ``Learn. Config.'' = learning configuration, ``TF'' = training-free, and ``Tr'' = trainable.}
\label{tab:profile_design_space_compact}
\vspace{-3mm}
\begin{tabular}{@{} l c c c c S[table-format=1.3] @{}}
\toprule
\textbf{Method} 
& \textbf{Org. Form} 
& \textbf{Rep. Type} 
& \textbf{Aggre. Hop} 
& \textbf{Learn. Config.} 
& {\textbf{Average Performance}} \\
\midrule
Oracle           & --         & --   & -- & -- & 0.679 \\
Single-Best      & --         & --   & -- & -- & 0.547 \\
Random Selection & --         & --   & -- & -- & 0.508 \\
\midrule
PlainText        & Flat       & Text & 0  & TF & 0.532 \\
\dashmidrule
EmbGNN-1hop      & Structured & Emb  & 1  & TF & 0.531 \\
EmbGNN-2hop      & Structured & Emb  & 2  & TF & 0.560 \\
EmbGNN-3hop      & Structured & Emb  & 3  & TF & 0.534 \\
EmbGNN-4hop      & Structured & Emb  & 4  & TF & 0.577 \\
\dashmidrule
TextGNN-1hop     & Structured & Text & 1  & TF & 0.526 \\
TextGNN-2hop     & Structured & Text & 2  & TF & 0.550 \\
TextGNN-3hop     & Structured & Text & 3  & TF & 0.566 \\
TextGNN-4hop     & Structured & Text & 4  & TF & 0.580 \\
\dashmidrule
TrainGNN-1hop    & Structured & Emb  & 1  & T  & \textbf{0.613} \\
TrainGNN-2hop    & Structured & Emb  & 2  & T  & \textbf{0.613} \\
TrainGNN-3hop    & Structured & Emb  & 3  & T  & 0.600 \\
TrainGNN-4hop    & Structured & Emb  & 4  & T  & 0.555 \\
\bottomrule
\end{tabular}
\end{table*}

\subsection{Profile Structure Improves Training-Free Cold-start Routing}

Table~\ref{tab:profile_design_space_compact} shows that profile structure substantially affects training-free cold-start routing. 
PlainText, which simply concatenates public signals, underperforms Single-Best (0.532 vs.\ 0.547), suggesting that flat profiles are insufficient. 
In contrast, structured profiles with sufficient aggregation depth close this gap: TextGNN-4hop (0.580) and EmbGNN-4hop (0.577) exceed Single-Best, showing that structural integration can improve routing without router training. 
Shallow aggregation is less effective, as TextGNN-1hop (0.526) and EmbGNN-1hop (0.531) provide little or no gain over PlainText, indicating that one-hop propagation offers insufficient context for distinguishing model capabilities. 
As depth increases, TextGNN improves steadily, whereas EmbGNN varies across depths, suggesting that aggregation depth interacts with representation type.

\subsection{Trainable Configurations Yield Larger but Depth-Sensitive Gains}

Trainable profiles further improve routing performance. 
TrainGNN-1hop and TrainGNN-2hop both reach 0.613, outperforming the best training-free profile (TextGNN-4hop, 0.580), and moving closer to Oracle performance (0.679). 
However, TrainGNN is sensitive to aggregation depth: performance drops from 0.613 at 2 hops to 0.600 at 3 hops and 0.555 at 4 hops. 
This degradation is consistent with over-smoothing, where repeated aggregation homogenizes node representations and weakens their discriminability.
By contrast, training-free profiles generally benefit from deeper aggregation. 

\begin{table*}[t]
\centering
\small
\caption{\textbf{Cold-start routing performance across public data sources.}
We compare benchmark-, domain-, and model-family-level information across three profile configurations. 
Method abbreviations follow Table~\ref{tab:profile_design_space_compact}. 
All configurations include model nodes; profile design details are provided in Section~\ref{section_exp_setup:design_choice}.}
\label{tab:source_ablation}
\vspace{-3mm}
\begin{tabular}{@{} l c c c S[table-format=1.3] @{}}
\toprule
\textbf{Method} 
& \textbf{Benchmark} 
& \textbf{Domain} 
& \textbf{Model Family} 
& {\textbf{Average Performance}} \\
\midrule
\multirow{4}{*}{EmbGNN-3hop}
& \cmark & \cmark & \cmark & 0.534 \\
& \cmark & \cmark &   & 0.503 \\
& \cmark &   & \cmark & 0.551 \\
& \cmark &   &   & 0.500 \\
\midrule
\multirow{4}{*}{TextGNN-3hop}
& \cmark & \cmark & \cmark & 0.566 \\
& \cmark & \cmark &   & 0.520 \\
& \cmark &   & \cmark & 0.552 \\
& \cmark &   &   & 0.509 \\
\midrule
\multirow{4}{*}{TrainGNN-3hop}
& \cmark & \cmark & \cmark & 0.600 \\
& \cmark & \cmark &   & 0.548 \\
& \cmark &   & \cmark & 0.519 \\
& \cmark &   &   & 0.502 \\
\bottomrule
\end{tabular}
\end{table*}

\subsection{Model Metadata is More Reliable than Benchmark Domain}

Table~\ref{tab:source_ablation} shows that model family metadata consistently improves routing across all profile designs. 
Adding family information to benchmark evaluation yields gains for EmbGNN-3hop (0.551 vs.\ 0.500), TextGNN-3hop (0.552 vs.\ 0.509), and TrainGNN-3hop (0.519 vs.\ 0.502). 
This consistency suggests that architectural lineage provides a stable structural prior for estimating model capabilities, even when no query-response-reward data is available. 
Such metadata is likely less sensitive to benchmark coverage variation and therefore serves as a robust public signal for cold-start profiling.
In contrast, domain information is more dependent on the profile design and learning configuration. 
When added to Benchmark+Model Family, domain information substantially improves TrainGNN-3hop (0.600 vs.\ 0.519), suggesting that trainable profiles can learn to exploit coarse domain structure. 
However, it degrades EmbGNN-3hop (0.534 vs.\ 0.551), indicating that training-free embedding profiles may be more sensitive to noisy aligned domain signals.

\begin{table*}[t]
\centering
\small
\caption{\textbf{New-LLM integration performance across profile designs.}
Method abbreviations follow Table~\ref{tab:profile_design_space_compact}. 
NCIR denotes the New-LLM Correct Integration Rate defined in Eq.~\ref{eq:cold-start-performance}.}
\label{tab:router-performance-ncir}
\vspace{-3mm}
\begin{tabular}{@{} l S[table-format=1.3] S[table-format=1.3] S[table-format=1.3] S[table-format=1.3] @{}}
\toprule
\multirow{2}{*}{\textbf{Method}}
& \multicolumn{2}{c}{\textbf{MLPRouter}}
& \multicolumn{2}{c}{\textbf{GraphRouter}} \\
\cmidrule(lr){2-3} \cmidrule(lr){4-5}
& {\textbf{Average Performance}}
& {\textbf{NCIR}}
& {\textbf{Average Performance}}
& {\textbf{NCIR}} \\
\midrule
PlainText      & 0.532 & 0.000 & 0.532 & 0.000 \\
\dashmidrule
EmbGNN-1hop    & 0.602 & \bfseries 0.283 & 0.527 & 0.198 \\
EmbGNN-2hop    & 0.612 & 0.020 & 0.547 & 0.079 \\
EmbGNN-3hop    & \bfseries 0.624 & 0.272 & \bfseries 0.613 & \bfseries 0.411 \\
\dashmidrule
TextGNN-1hop   & 0.610 & 0.007 & 0.541 & 0.000 \\
TextGNN-2hop   & 0.592 & 0.000 & 0.610 & 0.000 \\
TextGNN-3hop   & 0.594 & 0.007 & 0.503 & 0.000 \\
\dashmidrule
TrainGNN-1hop  & 0.530 & 0.000 & 0.582 & 0.400 \\
TrainGNN-2hop  & 0.529 & 0.000 & 0.524 & 0.000 \\
TrainGNN-3hop  & 0.530 & 0.000 & 0.515 & 0.013 \\
\bottomrule
\end{tabular}
\end{table*}

\subsection{Embedding-Based Profiles Enable New-LLM Integration}

Table~\ref{tab:router-performance-ncir} reveals a clear gap among profile types in routing queries to newly introduced LLMs. 
PlainText yields NCIR = 0.000 for both MLPRouter and GraphRouter, showing that flat profiles provide no effective signal for new-LLM integration. 
Text-based structured profiles also perform poorly on NCIR: TextGNN remains near zero across hops, reaching at most 0.007 with MLPRouter, despite competitive average performance. 
This gap indicates that textual profiles can support overall routing quality but fail to position the new LLM as a viable candidate within the router's decision space.

In contrast, embedding-based structured profiles consistently improve new-LLM integration. 
EmbGNN-3hop achieves NCIR = 0.411 with GraphRouter and 0.272 with MLPRouter, while also obtaining the best average performance for both routers (MLPRouter: 0.624; GraphRouter: 0.613). 
These results show that dense graph-based representations better transfer to unseen LLMs and can improve both new-model integration and overall routing quality.

\subsection{Profile--Router Co-Design Determines New-LLM Generalization}

New-LLM integration further depends on the alignment between profile design and router architecture. 
TrainGNN illustrates this dependence: TrainGNN-1hop achieves NCIR = 0.400 with GraphRouter but NCIR = 0.000 across all hops with MLPRouter under the same profile. 
This suggests that graph-structured routers are better able to exploit relational information encoded by trainable GNN profiles, whereas MLP-based routers struggle to use such structure for unseen models.
A similar pattern appears for EmbGNN. 
EmbGNN-3hop performs best with GraphRouter (NCIR = 0.411), whereas MLPRouter reaches its best NCIR at 1-hop (0.283), indicating that the optimal aggregation depth is router-dependent. 
Overall, GraphRouter achieves higher or equal NCIR across profile designs, suggesting that graph-structured routers are better aligned with relational public-signal profiles. 
Thus, effective new-LLM integration requires not only informative profiles, but also profile--router co-design.

\section{Conclusion}

In this work, we propose \method, a graph-based profiling framework for cold-start LLM routing. 
\method constructs model profiles from public signals in technical reports or model cards, including model family metadata, model descriptions, reported benchmark scores, and benchmark domain information, and organizes them into a heterogeneous graph. 
This enables systematic analysis of profile design along four dimensions: \textit{organizational form}, \textit{representation type}, \textit{aggregation depth}, and \textit{learning configuration}.
We evaluate \method in two settings: \textit{training-free cold-start routing} and \textit{new-LLM integration}. 
Our results reveal three key findings.
First, structured profiles substantially improve training-free cold-start routing. 
Second, model family metadata is more reliable than benchmark domain information across profile designs. 
Third, effective new-LLM integration depends critically on profile--router co-design. 
Overall, \method demonstrates that public-signal profiling provides a viable basis for cold-start routing and highlights profile design as an important direction for adapting routing systems to the expanding LLM ecosystem.

\newpage
\section*{Limitations}

This work has several limitations. 
First, public signals are incomplete and inconsistently reported across LLMs. 
Benchmark scores and model metadata are often drawn from technical reports or model cards with different evaluation protocols, benchmark coverage, and reporting granularity, which may introduce reporting bias into the constructed profiles.
Second, our experiments cover representative profile designs and router architectures, but do not exhaustively evaluate all possible routers, model families, or deployment scenarios. 
The effectiveness of profile--router co-design may therefore vary with the candidate model pool and routing architecture.
Finally, the new-LLM integration setting assumes that the router incorporates a newly released model using only its public profile and without retraining. 
In practice, routing systems may benefit from continually updating profiles as query-response-reward interactions accumulate. 
Exploring hybrid cold-start and continual adaptation strategies remains an important direction for future work.
\bibliography{custom}

\appendix

\section{Appendix}

\subsection{Implementation Details}

All experiments were conducted on 1 NVIDIA A6000 GPU. Each profile construction with the text-based GNN required approximately 0.2 GPU hours due to iterative LLM calls using vLLM. Each profile construction with the emb-based GNN required approximately 0.1 GPU hours in total. Each profile construction with the trainable GNN experiments required approximately 0.3 GPU hours in total. Inference outputs from the candidate LLMs were obtained via the NVIDIA API, requiring approximately 15 hours in total. We use the \texttt{allenai/longformer-base-4096} checkpoint from HuggingFace for encoding node text features, with a maximum sequence length of 4096 tokens. For the text-based GNN, Qwen3-8B \citep{yang2025qwen3} is queried with temperature 0 to ensure reproducibility. Graph neural network experiments are implemented using PyTorch Geometric (version 2.7.0). TrainGNN is trained for 100 epochs with a learning rate of 1e-3, batch size 64, and a masking ratio of 0.3 for both node and edge features, using the Adam optimizer. MLPRouter and GraphRouter are trained for 100 epochs with a learning rate of 1e-3 and 1e-4. Results in Tables 1–3 are reported from single runs. Random Selection is averaged over 6 random seeds (0–5).

\subsection{Data Sources for LLM Profile Construction}
\label{appendix:profile_evidence_description}

We describe the initial node features used to construct the interaction graph for LLM profiling, covering four types of nodes: model family, model, benchmark, domain, and query. All datasets and models used in this work are publicly available and used in accordance with their intended academic research purposes.

\subsubsection{Model Family Nodes}
Each model family node is initialized with a natural language description capturing its architectural design, training methodology, and general capabilities:
\begin{itemize}[noitemsep, topsep=4pt]
\item \textbf{Qwen2}: A family of decoder-only Transformer models developed by Alibaba Cloud, trained on large-scale multilingual corpora with improvements in data quality, alignment, and long-context handling.
\item \textbf{Gemma2}: An open model family released by Google, featuring grouped-query attention and interleaved local-global attention for efficient inference and high-quality language modeling.
\item \textbf{LLaMA}: A family of autoregressive Transformer models developed by Meta AI, widely adopted as foundation models for research and downstream applications, including instruction following and conversational agents.
\item \textbf{Mistral}: A family of high-efficiency decoder-only models developed by Mistral AI, incorporating grouped-query and sliding-window attention for scalable and memory-efficient inference.
\item \textbf{Mixtral}: A mixture-of-experts extension of the Mistral architecture that selectively activates sparse expert networks per token, achieving high model capacity with efficient computation.
\end{itemize}


\subsubsection{Model Nodes}
Each model node is initialized using its model family description as the base text feature, supplemented with model-specific attributes including parameter count, instruction-tuning status, and available model card information from representative model families such as Qwen~\citep{yang2025qwen3}, Gemma~\citep{team2024gemma}, Llama~\citep{grattafiori2024llama}, and Mixtral~\citep{jiang2024mixtral}. This allows model nodes to inherit shared architectural priors from their family while retaining individual characteristics.

\begin{table*}[!h]
\centering
\small
\caption{Model nodes and their descriptions used in the interaction graph.}
\label{tab:model_descriptions}
\begin{tabular}{lp{0.60\textwidth}}
\toprule
\textbf{Model} & \textbf{Description} \\
\midrule
\multicolumn{2}{l}{\textit{Candidate LLMs}} \\
\midrule
Qwen2.5-7B-Instruct & An upgraded 7B Qwen model with enhanced multilingual capabilities across diverse language tasks. \\
Gemma-2-9B-IT & A 9B instruction-tuned model from Google designed for general text processing and conversational applications. \\
Llama-3.1-8B-Instruct & Meta's 8B model from the Llama-3 series, designed for conversational AI and complex reasoning tasks. \\
Mixtral-8x7B-Instruct & A 56B mixture-of-experts model composed of eight 7B expert models, optimized for creative text generation. \\
Mixtral-8x22B-Instruct & An advanced 176B MoE model comprising eight 22B expert components, delivering strong performance across diverse tasks. \\
Llama-3.2-3B-Instruct & Meta's ultra-lightweight 3B model optimized for speed and efficiency, ideal for simple tasks requiring fast responses. \\
Mistral-Small-24B-Instruct & Mistral AI's latest compact model delivering strong performance from 24B parameters, excelling at instruction-following tasks. \\
\midrule
\multicolumn{2}{l}{\textit{Auxiliary Models}} \\
\midrule
Llama-3.3-70B-Instruct & Meta's 70B multilingual instruction model focused on high-quality dialogue, reasoning, coding, and tool use. \\
Qwen2.5-3B-Instruct & A 3B instruction model in the Qwen2.5 family, suited for low-cost applications and efficient local inference. \\
Qwen2.5-14B-Instruct & A 14B instruction model offering strong reasoning, knowledge use, and instruction-following for production workflows. \\
Qwen2.5-32B-Instruct & A 32B model built for stronger reasoning, richer world knowledge, and reliable long-form generation. \\
Qwen2.5-72B-Instruct & The 72B model in the Qwen2.5 series, built for top-tier reasoning and knowledge-intensive generation. \\
Gemma-2-2B-IT & Google's 2B instruction-tuned Gemma 2 model, offering a balanced blend of reasoning and response generation. \\
Gemma-2-27B-IT & Google's 27B instruction-tuned Gemma 2 model, delivering strong reasoning and response quality for high-quality workloads. \\
Llama-3.2-1B-Instruct & Meta's 1B instruction model optimized for fast, efficient text generation in constrained environments. \\
Mistral-Nemo-Instruct & A compact yet capable 12B instruction model jointly developed by Mistral AI and NVIDIA, strong in chat, coding, and multilingual tasks. \\
Qwen2.5-7B-Instruct-1M & Extended-context version of Qwen2.5-7B, supporting up to 1M tokens for long-document analysis and complex workflows. \\
Qwen2.5-14B-Instruct-1M & Combines stronger 14B reasoning with 1M token context support for advanced long-context enterprise workflows. \\
Qwen2-7B-Instruct & A 7B instruction model from Qwen2, offering a strong balance of chat quality, reasoning, and multilingual usability. \\
Qwen2-72B-Instruct & The 72B instruction model in the Qwen2 family, designed for premium assistants and demanding production workloads. \\
Llama-3.1-70B-Instruct & Meta's high-capability multilingual 70B instruction model for strong dialogue, reasoning, coding, and knowledge-intensive generation. \\
Ministral-8B-Instruct & Mistral AI's edge-focused 8B model, built for latency-sensitive assistants and compact production systems. \\
Mistral-Small-Instruct-2409 & A capable mid-sized 22B instruction model for general text generation, multilingual tasks, and function-calling workflows. \\
Mistral-Large-Instruct-2411 & Mistral AI's advanced 123B model built for state-of-the-art reasoning, coding, and long-context understanding. \\
\bottomrule
\end{tabular}
\end{table*}

\subsubsection{Benchmark Nodes}
Each benchmark node is initialized with a natural language description of the benchmark. Table~\ref{tab:dataset_descriptions} summarizes all datasets used in this work along with their descriptions.

\begin{table*}[!h]
\centering
\small
\caption{Benchmark nodes and their descriptions used in the interaction graph.}
\label{tab:dataset_descriptions}
\begin{tabular}{lp{10cm}}
\toprule
\textbf{Benchmark} & \textbf{Description} \\
\midrule
BBH~\cite{suzgun2023challenging} & A challenging subset of BIG-Bench focusing on tasks where earlier models performed significantly below human level, spanning multi-step arithmetic, logical reasoning, and complex language understanding. \\
MATH500~\cite{lightman2024let} & A curated 500-problem subset of the MATH benchmark covering competition-level mathematics including algebra, geometry, number theory, and combinatorics. \\
GPQA~\cite{rein2023gpqa} & A graduate-level benchmark with expert-authored multiple-choice questions in physics, chemistry, and biology, designed to resist simple retrieval-based answering. \\
MuSR~\cite{sprague2024musr} & A benchmark for multi-step and structured reasoning that requires integrating multiple pieces of information through sequential inference. \\
MMLU-Pro~\cite{wang2024mmlu} & An enhanced version of MMLU with more difficult questions and expanded answer choices, designed to better evaluate reasoning ability. \\
MMLU~\cite{hendrycks2020measuring} & A broad multiple-choice benchmark covering 57 academic and professional subjects across humanities, social science, STEM, and other domains. \\
C-Eval~\cite{huang2023c} & A Chinese standardized-exam benchmark spanning dozens of disciplines for evaluating Chinese language understanding and reasoning in exam-style settings. \\
AGIEval~\cite{zhong2024agieval} & A human-centric benchmark derived from official admission and qualification exams (e.g., SAT, Gaokao) to evaluate general reasoning and problem-solving. \\
TriviaQA~\cite{joshi2017triviaqa} & A large-scale QA benchmark with trivia questions and evidence documents, testing knowledge retrieval and answer generation under noisy real-world evidence. \\
Natural Questions~\cite{kwiatkowski2019natural} & A real-user QA benchmark based on anonymized Google queries with Wikipedia evidence, evaluating short-answer and long-answer question answering. \\
SQuAD~\cite{rajpurkar2016squad} & A reading comprehension benchmark of crowd-authored questions on Wikipedia passages, where answers are extracted text spans. \\
TheoremQA~\cite{chen2023theoremqa} & A STEM theorem-driven QA benchmark with university-level problems across math, CS/EE, physics, and finance, testing formal reasoning and theorem application. \\
CommonsenseQA~\cite{talmor2019commonsenseqa} & A multiple-choice commonsense benchmark built from ConceptNet relations, requiring implicit everyday knowledge. \\
WinoGrande~\cite{sakaguchi2021winogrande} & A large-scale pronoun coreference benchmark testing robust commonsense reasoning in Winograd-style disambiguation. \\
ARC-Challenge~\cite{clark2018think} & The difficult split of the AI2 Reasoning Challenge, containing grade-school science questions that require deeper reasoning beyond retrieval. \\
OpenBookQA~\cite{mihaylov2018can} & A science QA benchmark requiring multi-hop reasoning by combining core facts with external commonsense knowledge. \\
BoolQ~\cite{clark2019boolq} & A yes/no QA benchmark built from real user queries paired with evidence passages, testing binary reading comprehension and inference. \\
DROP~\cite{dua2019drop} & A reading comprehension benchmark requiring discrete reasoning such as counting, comparison, and arithmetic over paragraphs. \\
GSM8K~\cite{cobbe2021training} & A grade-school math word-problem benchmark with multi-step natural-language solutions for evaluating arithmetic reasoning. \\
MGSM~\cite{shi2022language} & A multilingual extension of GSM8K-style math problems enabling cross-lingual evaluation of multi-step mathematical reasoning. \\
HumanEval~\cite{chen2021evaluating} & A code-generation benchmark of hand-written programming problems with hidden unit tests evaluating functional correctness. \\
MBPP~\cite{austin2021program} & A benchmark of around one thousand crowd-sourced entry-level Python tasks with reference tests for practical code generation. \\
TruthfulQA~\cite{lin2022truthfulqa} & A benchmark measuring whether models produce truthful answers rather than imitating common human misconceptions. \\
\bottomrule
\end{tabular}
\end{table*}

\subsubsection{Domain Nodes}
Each domain node represents a high-level capability category. We define six domains, each characterized by a natural language description that serves as the initial text feature:
\begin{itemize}[noitemsep, topsep=4pt]
\item \textbf{Knowledge}: Tasks requiring broad factual knowledge, academic understanding, and evidence-grounded question answering across domains.
\item \textbf{Reasoning}: Tasks requiring commonsense reasoning, multi-step inference, logical deduction, and robust decision making.
\item \textbf{QA}: Tasks centered on question answering with retrieval, reading comprehension, and answer faithfulness to provided evidence.
\item \textbf{Math}: Tasks requiring arithmetic, symbolic manipulation, theorem application, and multi-step mathematical problem solving.
\item \textbf{Coding}: Tasks requiring program synthesis, debugging, and functional correctness under executable unit tests.
\item \textbf{Alignment}: Tasks evaluating instruction following, helpfulness, harmlessness, truthfulness, and preference alignment in assistant behavior.
\end{itemize}

\subsubsection{Query Nodes}

Query nodes represent individual queries sampled from each benchmark dataset. For each dataset, up to 1,000 queries are randomly sampled to serve as query nodes in the interaction graph. Each query node is initialized by encoding the raw query text using a pre-trained language model. Table~\ref{tab:query_datasets} lists the datasets and their corresponding Hugging Face identifiers used for query sampling.

\begin{table*}[t]
\centering
\small
\caption{Hugging Face dataset identifiers used for query node construction.}
\label{tab:query_datasets}
\begin{tabular}{ll}
\toprule
\textbf{Dataset} & \textbf{Hugging Face Identifier} \\
\midrule
IFEval & \texttt{google/IFEval} \\
BBH & \texttt{lukaemon/bbh} \\
MATH500 & \texttt{HuggingFaceH4/MATH-500} \\
GPQA & \texttt{Idavidrein/gpqa} \\
MuSR & \texttt{TAUR-Lab/MuSR} \\
MMLU-Pro & \texttt{TIGER-Lab/MMLU-Pro} \\
EvalPlus & \texttt{evalplus/humanevalplus} \\
MultiPL-E & \texttt{nuprl/MultiPL-E} \\
C-Eval & \texttt{ceval/ceval-exam} \\
AGIEval English & \texttt{lighteval/agi\_eval\_en} \\
SQuAD & \texttt{rajpurkar/squad} \\
TheoremQA & \texttt{TIGER-Lab/TheoremQA} \\
WinoGrande & \texttt{allenai/winogrande} \\
BoolQ & \texttt{google/boolq} \\
DROP & \texttt{ucinlp/drop} \\
TruthfulQA & \texttt{domenicrosati/TruthfulQA} \\
WildBench & \texttt{allenai/WildBench} \\
\bottomrule
\end{tabular}
\end{table*}

\subsection{Prompts for Text-based GNN}
\label{appendix:prompts_for_textgnn}

At each propagation hop, every node in the interaction graph is updated by an LLM that synthesises information from its local neighbourhood. The prompts are designed to reflect the heterogeneous structure of the graph, with distinct templates for each node type.

\begin{table*}[!h]
\centering
\small
\caption{Prompt templates for Text-GNN aggregation across different node types.}
\label{tab:prompt_templates}
\begin{tabular}{lp{0.25\textwidth}p{0.25\textwidth}p{0.25\textwidth}}
\toprule
\textbf{Node Type} & \textbf{Input Context} & \textbf{Instruction} & \textbf{Output} \\
\midrule
Model &
Model family; benchmark scores grouped by domain; representative queries ranked by similarity to connected datasets &
Synthesise all context into a unified capability profile covering architecture, domain-level performance, and query suitability &
3--5 sentence capability profile \\
\addlinespace
Dataset &
Parent domain; models evaluated with scores; representative queries from the benchmark &
Describe what capability the benchmark evaluates, which models perform well or poorly, and what query types it covers &
2--4 sentence benchmark profile \\
\addlinespace
Domain &
All benchmark datasets belonging to this domain &
Characterise the capability area and summarise the benchmark landscape within it &
2--4 sentence domain profile \\
\addlinespace
Model Family &
All model nodes that instantiate this architecture &
Describe key design characteristics and the typical capability profile of models built on this architecture &
2--4 sentence architecture profile \\
\addlinespace
Query &
\multicolumn{3}{l}{\textit{Not updated — raw query text is preserved throughout all hops as a stable semantic anchor.}} \\
\bottomrule
\end{tabular}
\end{table*}

\subsection{Dataset Statistics}
\label{appendix:dataset}

Table~\ref{tab:datasets} summarizes the datasets used in this work, divided into two groups: those used for evidence graph construction during LLM profiling, and those used for routing evaluation.

\begin{table*}[t]
\centering
\caption{Overview of Datasets for Profile Construction and Routing Evaluation.}
\label{tab:datasets}
\resizebox{0.95\textwidth}{!}{
\begin{tabular}{l l c c c}
\toprule
\textbf{Usage} & \textbf{Dataset} & \textbf{Benchmark Type} & \textbf{Metric} & \textbf{Cases} \\
\midrule
\multirow{15}{*}{Profile Construction}
 & BBH          & Reasoning             & Accuracy & 1000 \\
 & MATH500      & Math                  & Accuracy & 500  \\
 & GPQA-Diamond & Knowledge / Reasoning & Accuracy & 198  \\
 & MUSR         & Reasoning             & Accuracy & 756  \\
 & MMLU-Pro     & Knowledge             & Accuracy & 1000 \\
 & AGIEval      & Knowledge             & Accuracy & 29   \\
 & TheoremQA    & Math / Reasoning      & Accuracy & 800  \\
 & DROP         & Reasoning             & Accuracy & 1000 \\
 & TruthfulQA   & Reasoning             & Accuracy & 817  \\
 & WinoGrande   & Reasoning             & Accuracy & 1000 \\
 & BoolQ        & Reasoning             & Accuracy & 1000 \\
 & C-Eval       & Knowledge             & Accuracy & 1000 \\
 & SQuAD        & Knowledge             & Accuracy & 1000 \\
 & MultiPL-E    & Coding                & Accuracy & 1000 \\
 & EvalPlus     & Coding                & Accuracy & 164  \\
\midrule
\multirow{12}{*}{Routing Evaluation}
 & MGSM         & Math                  & Accuracy & 50 \\
 & GSM8K        & Math                  & Accuracy & 50 \\
 & AgentVerse   & Reasoning             & Accuracy & 50 \\
 & CommonsenseQA & Reasoning            & Accuracy & 50 \\
 & OpenBookQA   & Reasoning             & Accuracy & 50 \\
 & ARC-Challenge & Reasoning            & Accuracy & 50 \\
 & MMLU         & Knowledge             & Accuracy & 50 \\
 & NaturalQA    & Knowledge             & Accuracy & 50 \\
 & TriviaQA     & Knowledge             & Accuracy & 50 \\
 & CommonGen    & Knowledge             & Accuracy & 50 \\
 & MBPP         & Coding                & Accuracy & 50 \\
 & HumanEval    & Coding                & Accuracy & 50 \\
\bottomrule
\end{tabular}
}
\end{table*}

\subsection{LLM Statistics}
\label{appendix:llm}

Table~\ref{tab:llms} summarizes the LLMs used in this work, divided into candidate models that participate in routing and auxiliary models that serve as additional graph context nodes during profile construction.

\begin{table*}[t]
\centering
\caption{Statistics of Candidate and Auxiliary LLMs.}
\label{tab:llms}
\resizebox{0.75\textwidth}{!}{
\begin{tabular}{llcc}
\toprule
\textbf{Role} & \textbf{LLM} & \textbf{Size} & \textbf{Model Family} \\
\midrule
\multirow{8}{*}{Candidate}
 & Llama-3.2-3B-Instruct           & 3B   & Llama   \\
 & Qwen2.5-7B-Instruct             & 7B   & Qwen2.5 \\
 & Llama-3.1-8B-Instruct           & 8B   & Llama   \\
 & Gemma-2-9B-IT                   & 9B   & Gemma2  \\
 & Mistral-Small-24B-Instruct-2501 & 24B  & Mistral \\
 & Mixtral-8x7B-Instruct-v0.1      & 56B  & Mixtral \\
 & Llama-3.3-70B-Instruct          & 70B  & Llama   \\
 & Mixtral-8x22B-Instruct-v0.1     & 176B & Mixtral \\
\midrule
\multirow{14}{*}{Auxiliary}
 & Llama-3.2-1B-Instruct           & 1B   & Llama   \\
 & Gemma-2-2B-IT                   & 2B   & Gemma2  \\
 & Qwen2.5-3B-Instruct             & 3B   & Qwen2.5 \\
 & Qwen2-7B-Instruct               & 7B   & Qwen2   \\
 & Qwen2.5-7B-Instruct-1M          & 7B   & Qwen2.5 \\
 & Ministral-8B-Instruct-2410      & 8B   & Mistral \\
 & Mistral-Nemo-Instruct-2407      & 12B  & Mistral \\
 & Qwen2.5-14B-Instruct            & 14B  & Qwen2.5 \\
 & Qwen2.5-14B-Instruct-1M         & 14B  & Qwen2.5 \\
 & Mistral-Small-Instruct-2409     & 22B  & Mistral \\
 & Gemma-2-27B-IT                  & 27B  & Gemma2  \\
 & Qwen2.5-32B-Instruct            & 32B  & Qwen2.5 \\
 & Qwen2-72B-Instruct              & 72B  & Qwen2   \\
 & Qwen2.5-72B-Instruct            & 72B  & Qwen2.5 \\
 & Llama-3.1-70B-Instruct          & 70B  & Llama   \\
 & Mistral-Large-Instruct-2411     & 123B & Mistral \\
\bottomrule
\end{tabular}
}
\end{table*}

\subsection{AI Usage}

We used AI-assisted writing tools (ChatGPT, Claude) to polish the language and improve the clarity of this paper. All technical content, experimental design, analysis, and conclusions are entirely the work of the authors. The use of these tools is limited to linguistic refinement and does not affect the scientific contributions of this work.

\end{document}